\documentclass[amsmath,amssymb,reprint,groupedaddress]{revtex4-1}

\pdfoutput=1

\usepackage{graphicx}
\usepackage{dcolumn}
\usepackage{bm}
\usepackage{siunitx}
\usepackage{lipsum}
\usepackage{enumitem}
\usepackage{relsize}
\usepackage{threeparttable}
\usepackage{multirow}

\begin{document}
\title{
Free Energy Calculations of Membrane Permeation: \\
Challenges due to Strong Headgroup--Solute Interactions
}
\author{Nihit Pokhrel}
\author{Lutz Maibaum}
\affiliation{Department of Chemistry, University of Washington, Seattle, WA 98195 }
\begin{abstract}
Understanding how different classes of molecules move across biological membranes is a prerequisite to predicting a solute's permeation rate, which is a critical factor in the fields of drug design and pharmacology. We use biased Molecular Dynamics computer simulations to calculate and compare the free energy profiles of translocation of several small molecules across 1,2-dioleoyl-sn-glycero-3-phosphocholine (DOPC) lipid bilayers as a first step towards determining the most efficient method for free energy calculations. We study the translocation of arginine, a sodium ion, alanine, and a single water molecule using the Metadynamics, Umbrella Sampling, and Replica Exchange Umbrella Sampling techniques. Within the fixed lengths of our simulations, we find that all methods produce similar results for charge-neutral permeants, but not for polar or positively charged molecules. We identify the long relaxation timescale of electrostatic interactions between lipid headgroups and the solute to be the principal cause of this difference, and show that this slow process can lead to an erroneous dependence of computed free energy profiles on the initial system configuration. We demonstrate the use of committor analysis to validate the proper sampling of the presumed transition state, which in our simulations is achieved only in replica exchange calculations. Based on these results we provide some useful guidance to perform and evaluate free energy calculations of membrane permeation.
\end{abstract}

\pacs{Valid PACS appear here}
\maketitle

\section{\label{sec:intro}Introduction}

Cell membranes are impermeable to many ions and hydrophilic compounds. The tight packing of phospholipids, sterols, and other membrane components together with the hydrophobic interior of the membrane prevent entry of foreign particles into the cell. There are, however, several important exceptions to this general rule, including larger molecules such as cell penetrating peptides (CPPs) that are capable of permeating the bilayer. Understanding the mechanism and kinetics of membrane permeation by these molecules facilitates the understanding of pore formation \cite{Herce2007, Hu2016}, passive translocation of solutes \cite{Tieleman2006, Piantavigna2015}, and drug delivery across cell membranes \cite{Lindgren2000}. 
As a consequence, there is an enormous interest in discovering new biomolecules that can spontaneously diffuse across lipid membranes. 

One characteristic common to most membrane-permeating solutes is that they are positively charged \cite{Futaki2002}, and CPPs fall into this category. They have gained much attention due to their ability to deliver large molecular cargoes into the interior of a cell~\cite{Zorko2005}. In addition to CPPs, peptides with positively charged amino acid residues like arginine and lysine have also been studied in an attempt to use them as drug carriers across lipid bilayers~\cite{Frankel1988}. Despite the massive interest in these systems, the precise interactions of these carrier peptides with the membrane still remains unclear~\cite{Deshayes2005, Bechara2013, Milletti2012}. In order to design novel drug carriers, a detailed understanding of how small solutes such as individual amino acids move across lipid bilayers is necessary. This will allow to predict which peptides can act as vehicles for drug delivery across cell membranes. 

Membrane partitioning is difficult to measure experimentally because of the complexity of lipid bilayer systems~\cite{Clarke2001, Swift2013, Cardenas2015}. Computational approaches such as Molecular Dynamics (MD) simulations can provide atomistic-level insights into interactions of amino acids with lipid bilayers, complementing experimental studies~\cite{Schutz2000, Ciobanasu2010}. Free energy profiles are often calculated as a first step towards understanding the permeation mechanism and predicting the kinetics of translocation \cite{Johansson2009, Dorairaj2007, MacCallum2008, Wennberg2012, Tieleman2006a, Dickson2017, Lee2016}. Conventional MD is not well suited for free energy calculations of rare events that involve high energy barriers, which include membrane translocation. These large barriers restrict the simulation from efficiently sampling the entire configuration space because the system remains trapped in a local free energy minimum. A number of accelerated sampling methods that bias the system to sample otherwise inaccessible regions have been developed. These methods aid in correctly predicting the free energy of many interesting biological processes. While all these methods improve the statistical sampling and reduce the amount of required computational resources, it is often not apparent which method is best suited for a specific system of interest. In the last five years considerable attention has been given to identifying the sources of sampling errors, where the difficulty lies in evaluating the convergence of results and estimating the associated uncertainty within these simulation techniques. Neale and coworkers have extensively studied convergence of free energy profiles of translocation of small solutes across lipid bilayers, specifically arginine translocation across DOPC~\cite{Neale2011, Neale2013, Neale2016}. They identified hidden barriers that depend on the interactions of the solute with water and lipid molecules.
Considering the growing interest of quantifying uncertainty in free energy calculations, we here compare the efficiency of three popular acceleration methods in the context of passive membrane translocation: well tempered metadynamics (WT-metaD) \cite{Barducci2008}, umbrella sampling (US) \citep{Torrie1977}, and replica exchange umbrella sampling (umbrella exchange, UE) \citep{Sugita1999}. Other methods that are frequently used in this context include Adaptive Biasing Force \citep{Darve2001} and Thermodynamic Integration \citep{Kirkwood1935}, but are not considered here. To compare these three methods, we use them to compute the free energy of translocation of arginine, a single sodium ion, different forms of alanine, and water across a DOPC bilayer. It should be noted that the translocation mechanism across a pure phospholipid bilayer might be different from that across a biological membrane, where channel proteins can facilitate the process~\citep{Johansson2009a}. However, studying small solutes traversing across such model membranes can provide valuable insights into general principles of passive membrane translocation. In addition, the solute molecules  were chosen because they are well studied in the literature \citep{Dorairaj2007, MacCallum2011,  Gumbart2011, Wilson1996, Sapay2009, Orsi, Comer2014} and represent a class of compounds that commonly appear in translocation problems and span a range of sizes, shapes and hydrophobicities \citep{Wimley1996}. An extensive list of translocation research can be found in recent review papers ~\citep{Awoonor-williams2016, Neale2016}. While most previous works focus on the mechanistic details of the permeation phenomenon itself, we concentrate on identifying and diagnosing generic convergence issues. In particular, we compute the committor distribution function to check if our simulations accurately sample the transition state ensemble (TSE) of the translocation process. We find that free energy calculations of positive and polar solutes do not converge within typical timescales of membrane simulations. The difficulty in relaxation of electrostatic interactions between the solute and lipid headgroups gives rise to hysteresis-like behavior, which decays only after extensive UE equilibration. Based on our results, we identify some useful diagnostic tools to evaluate the accuracy of calculated free energy profiles of membrane permeation. 

\section{\label{sec:methods}Methods}

\subsection{System Preparation}

We performed all simulations using Gromacs 4.6.7 \citep{VanDerSpoel2005} with the Plumed 2.1 plugin \citep{Bonomi2009} under periodic boundary conditions. Temperature and pressure were maintained at 320 K and 1 atm using the Nose-Hoover thermostat and Berendsen barostat, respectively. Long range electrostatic interactions were computed using the fourth order PME method \cite{Darden1993} with a Fourier spacing of 0.12 nm. The real space coulombic interaction was calculated up to 1.0 nm. Van der Waals interactions were calculated using a cutoff of 1.0 nm. Bond lengths within the solutes and lipids were constrained using the LINCS algorithm \citep{Hess1997}. 
 
The 1,2-dioleoyl-sn-glycero-3-phosphocholine (DOPC) bilayer was constructed using the united-atom Berger forcefield \citep{Berger1997} such that each monolayer consisted of 64 lipids. Water molecules were treated using the rigid simple point charge (SPC) model~\citep{Berendsen1981}. Each bilayer-water system was equilibrated for 10 ns before adding any permeant to the system. The permeants were modeled using the all-atom OPLS-AA forcefield \citep{Kaminski2001}. Cationic arginine with charged termini (NH$_3^+$ and COO$^-$) was constructed using the pdb2gmx tool of Gromacs. A system containing a single sodium ion was built using the genion tool of Gromacs. Three different forms of alanine were constructed: the first by truncating the side chain at the $\beta$-carbon with the $\alpha$-carbon replaced by a hydrogen; we call this form the side chain analog, where the alanine residue essentially becomes a methane molecule. This method of truncating amino acids has been used in the past to study amino acid interactions with model bilayer systems \citep{Johansson2008, Johansson2009}. The second form of alanine was constructed with neutral termini (NH$_2$ and COOH). We made the third form with charged termini (NH$_3^+$ and COO$^-$), a charge neutral but zwitterionic molecule. Thus we have studied the following permeants: 
\begin{itemize}[noitemsep]
	\item[i.] Arginine 
	\item[ii.] Sodium ion
	\item[iii.] Side chain analog of alanine 
	\item[iv.] Alanine with neutral termini 
	\item[v.] Zwitterionic alanine 
	\item[vi.] Water 
\end{itemize}
For the simulations of positively charged solutes a single chloride ion was added to achieve overall charge neutrality. Each permeant/water system was equilibrated for 10 ns. The equilibrated water/bilayer and water/permeant systems were then combined to form the final water/bilayer/permeant system, which was again equilibrated for 10 ns before any production run under NPT conditions. The Visual Molecular Dynamics (VMD) software was used to monitor and visually inspect all simulation trajectories \citep{Humphrey1996}.  
 
\subsection{Well Tempered Metadynamics (WT-metaD)}

Metadynamics is a biasing technique that overcomes sampling problems by adding a history-dependent bias potential $V_G(z,t)$ to the collective variable $z$, which itself is a function of the positions of the atoms in the simulation \citep{Laio2008}. The bias potential is constructed of Gaussian-shaped energy hills that are deposited along the collective variable,
\begin{equation}
V_G(z,t) = \omega_0 \mathlarger{\mathlarger{\sum}}_{t'<t}\exp\left(- \frac{(z-z(t'))^2}{2\sigma^2}\right),
\end{equation}
where  $\omega_0$ is the height of the added Gaussian, $\sigma$ is its width, and $z(t ')$ is the value of the collective variable at time $t'$. Each hill is deposited at a predefined rate and centered at a previously explored configurations, biasing the system towards configurations that have not yet been explored. At longer times, the sum of added hills can be used to calculate the unbiased free energy profile of the variable $z$. Well-tempered metadynamics (WT-metaD) is an improved form of metadynamics where the height $\omega_0$ of the hills decreases in previously visited regions, which guarantees correct convergence of the free energy profiles \citep{Barducci2008, Voth2014}. The choice of the metadynamics parameters is crucial for successful convergence \cite{Bochicchio2015}. The values we use for this work can be found in Table S1.

We choose the normal component of the distance vector between the center of mass of the solute and the center of mass of the lipid bilayer as the collective variable $z$. We used snapshots from a WT-metaD trajectory, with permeants at various positions relative to the bilayer, as the initial configurations for both US and UE calculations; the snapshots were taken when the solute first reached the desired distance after a minimum of 100 ns of WT-metaD simulation time.  

\subsection{Umbrella Sampling (US)}

Umbrella Sampling also adds a biasing potential to the system's Hamiltonian to enhance the sampling of configurations that are high in free energy \citep{Torrie1977}. In this case the biasing potential is static. We choose a sequence of ``windows'' that span the range of interest of the collective variable $z$. In the $i^{th}$ window the system is biased to remain close to a predetermined value $z_i$ by using a harmonic umbrella potential
\begin{equation}
V_i(z)=\frac{1}{2} k(z-z_i)^2,
\end{equation}
where $k$ is the stiffness of the harmonic potential. The results of $N$ independent simulations, each performed with a different value of $z_i$, are then combined using the weighted histogram analysis method (WHAM) to obtain an estimate of the unbiased free energy profile $F(z)$ \citep{Kumar1992}. Uncertainties were estimated using bootstrapping analysis as implemented in the g$\_$wham tool of the Gromacs simulation suite~\citep{Hub2010}.

We use the same reaction coordinate $z$ for the US calculation as for WT-metaD. For each solute we used a series of windows with spacing of 0.1 nm spanning the entire bilayer. A harmonic potential of $k=1000$ kJ/mol/nm$^{2}$ was used. 40 windows were used for arginine simulation and each window was run for 100 ns, resulting in 4.0 $\mu$s of total simulation time. 50 windows were constructed for all three forms of alanine. For the side chain analog and alanine with neutral termini, each window was simulated for 20 ns totaling 1 $\mu$s of simulation time. Zwitterionic alanine was simulated for 2.75 $\mu$s where each window was run for 55 ns (see table S2). 

\subsection{Replica Exchange Umbrella Sampling (UE)}

UE is very similar to US, with the addition that neighboring windows can exchange their configurations. In UE, $N$ parallel and independent simulations of the same system, biased at different values of $z_i$, are run, each of these $N$ simulations is called a replica. An exchange of configurations between neighboring replicas is attempted at a pre-determined frequency, and is accepted or rejected based on the Metropolis criterion. This technique is similar to other Replica Exchange schemes where different replicas are simulated at different temperatures, which improves the sampling at low temperatures by incorporating enhanced sampling at higher temperatures\citep{Hukushima1996}. Free energy profiles and the associated uncertainties were again computed using the g$\_$wham tool.

To allow for accurate comparisons between the simulation methods we use the exact same parameters for US and UE. Each replica was run for 165 ns for arginine and for 55 ns for alanine. For sodium and water, each replica was simulated for 27 ns and 8 ns, respectively. An exchange was attempted every 2 ps for all UE calculations. Table S2 contains information about simulation lengths for each solute using each of the three methods.   

Hydrophilic solutes are most likely found either in bulk water or at the water/membrane interface, which corresponds to a local minimum in the free energy profile $F(z)$. Because the membrane is symmetric in our simulations, there are two degenerate minima corresponding to the upper (A) and lower (B) half-spaces that are separated by the membrane. For a given configuration we define the committor $p_\text{B}$ as the probability that a trajectory initiated from this configuration with random initial velocities will reach the free energy minimum B on the lower ($z<0$) side of the membrane before it reaches minimum A on the upper side ($z>0$). The collection of configurations with $p_\text{B} = 1/2$ form the transition state surface that separates states that are more likely to go to A from those that are more likely to go to B. This definition of the transition surface does not necessarily correspond to a saddle point in a free energy landscape. It is preferred because of its intuitive kinetic interpretation and because it does not require a choice of a reaction coordinate in the transition region ~\cite{Bolhuis2002, Dellago2001}; instead it requires only a well-defined reactant and product state. In this picture a transition state can also be a metastable intermediate. 

Given the symmetric nature of our membrane, one might expect configurations in which a hydrophilic permeant is at the center of the bilayer to be transition states. To test this hypothesis we calculate the distribution of committors, $P(p_\text{B})$, over multiple states with $z=0$. If all configurations in this ensemble are indeed transition states, then $P(p_\text{B})$ will be sharply peaked at $p_\text{B} = 1/2$ \citep{Bolhuis2002}.

To compute the committor distribution function $P(p_\text{B})$ for an ensemble of configurations with $z=0$ as generated by the WT-metaD, US, and UE methods we take six such configurations from each of the three methods, and initiate four unbiased trajectories with random initial velocities from each of these eighteen configurations. We then count in how many of these four trajectories the solute reaches the lower ($z<0$) membrane/water interface before the upper interface. The resulting fraction serves as an estimate for the  $p_\text{B}$-value of a configuration, and we construct a histogram of these values over the six configurations obtained from each sampling method. These histograms are shown in Fig.~\ref{fig:committordistributions}.

Both the number of configurations sampled from the $z=0$ ensemble and the number of trajectories initiated from these configurations to estimate the committor are quite small. One typically needs better sampling to obtain accurate estimates of $p_\text{B}$ and $P(p_\text{B})$~\cite{Dellago2001}. However, the long timescale of solute motion across the membrane limits us to such relatively small numbers.

\section{\label{sec:results}Results}

\begin{figure}[t]
\includegraphics[width=\columnwidth]{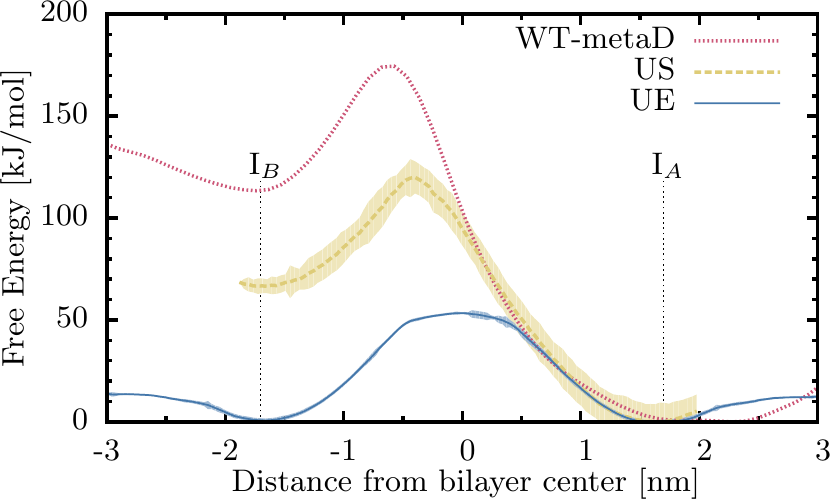}
\caption{\label{fig:Farginine}Free Energy of arginine translocation as a function of distance from the bilayer center using WT-metaD, US and UE with shaded error bars for US and UE. All three profiles have two minima: one at the interface of upper leaflet and water (I$_\text{A}$) and one at the interface of the lower leaflet and water (I$_\text{B}$). The three lines differ: profiles generated using WT-metaD and US are asymmetric and the maximum does not occur at $z=0$. The UE profile, obtained after extensive equilibration, is symmetric and has its maximum at $z= 0$. }
\end{figure}

\begin{figure}[t]
\includegraphics[width=\columnwidth]{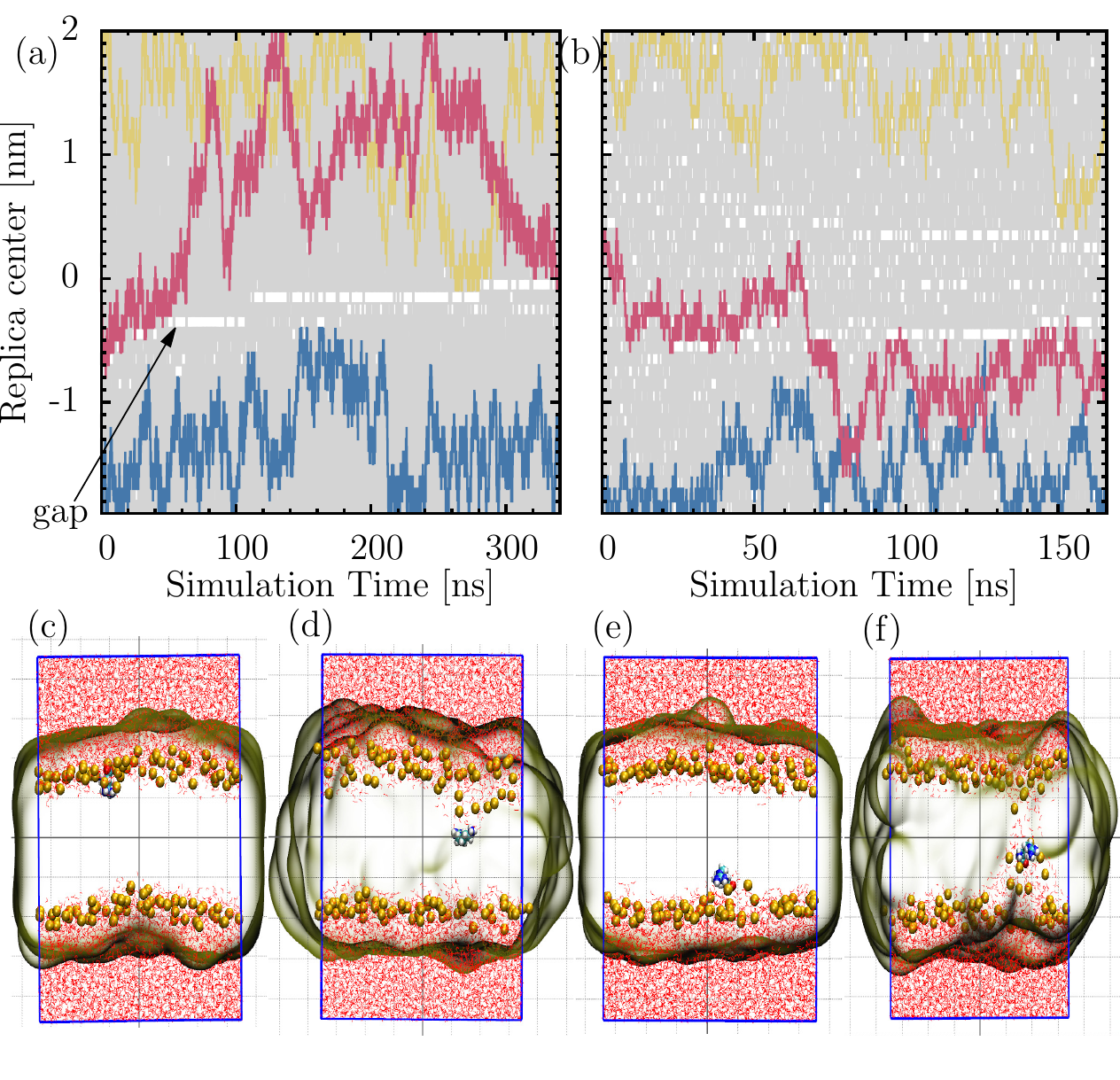}
\caption{\label{fig:UEarginine} UE results for arginine. (a) Exchange pattern lines for 40 replicas. Replicas starting at $z=$ 2.0 nm, 0 nm and -1.9 nm are highlighted. Initial configurations for these replicas are shown in ((c)-(e)) with water shown in red, headgroup phosphates in orange, membrane outline in dark yellow, and the solute at various position across the DOPC bilayer. A gap in the exchange pattern (indicated by the arrow) is clearly visible. Using the final configuration of the $z=0$ replica (shown in (f)) as the initial condition for the center replicas in a new UE simulation (b), this gap is no longer present, indicating even exchanges between all 40 replicas.}
\end{figure}

\subsection{Committor Distribution Function}
\begin{figure}[bt]
\includegraphics[width=\columnwidth]{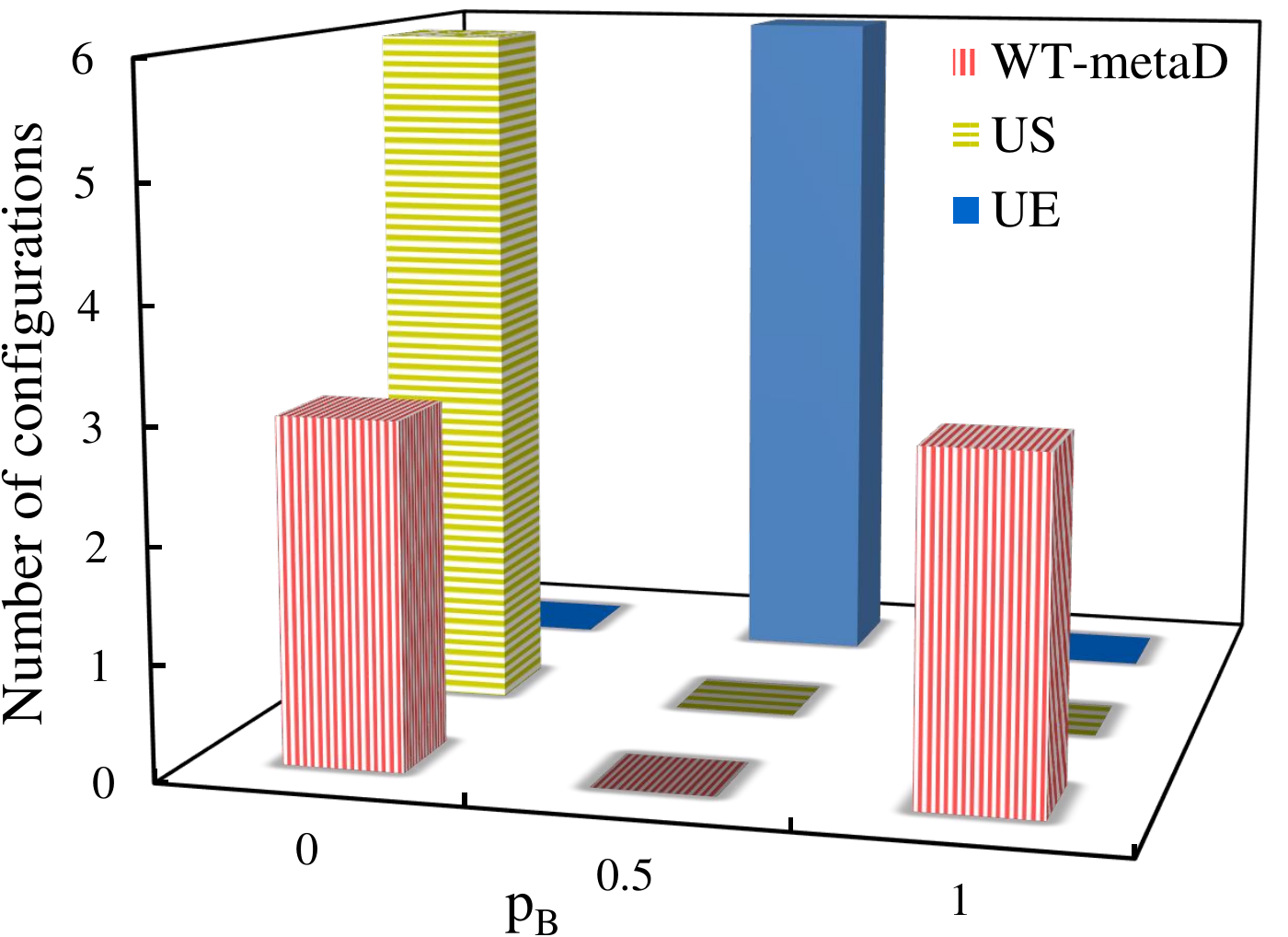}
\caption{\label{fig:committordistributions} Committor histogram for 3 ensembles with arginine at $z=0$ nm. The single peak at $p_B=0.5$ for UE shows that arginine has no preference for either side of the membrane in all six tested configurations. Peaks at $p_\text{B} = 0$ and $p_\text{B} = 1$ for WT-metaD and US show that each of the tested configurations has a strong bias towards one side of the membrane.
}
\end{figure}

We compare the free energy profiles of translocation  for arginine, a single sodium ion, alanine, and a water molecule as calculated by the WT-metaD, US and UE methods. When calculating such profiles for symmetric membranes, it is tempting to take advantage of this symmetry and to calculate the free energy only for one half-space, either $z \geq 0$ or $z \leq 0$, and obtain the other half by simply mirroring the result with respect to the $z=0$ axis. Another way to exploit this symmetry is to compute the free energy profile for the whole range of $z$, and then average together the free energies above and below the membrane center. Both approaches are formally correct and yield symmetric free energy profiles by construction. However, we will see that they can conceal signatures of poor convergence of the simulations. We therefore choose to compute free energy profiles for the entire range of the collective variable $z$. This prevents us from losing any information regarding the translocation process and also helps us evaluate the discrepancy, if any, between the free energy profiles calculated using a single versus both monolayers, as suggested previously \citep{Neale2013, Lee2016}.  

\subsection{Arginine}
 
In Fig.~\ref{fig:Farginine} we show the free energy  of arginine translocation across a DOPC bilayer as a function of arginine distance from the membrane center as computed by WT-metaD, US, and UE. Each profile is shifted vertically to have the global minimum at zero. We see that all three profiles have local minima at the interface of the bilayer and water ($z \approx \pm$ 1.7 nm), which indicates that arginine prefers the interface over both the hydrophobic core and the bulk water. At these minima, arginine forms strong electrostatic interactions with lipid headgroups and remains solvated by water.

Other than the position of these local minima, there are stark differences between the three profiles. This is unexpected as all three methods in principle converge to the same free energy function. Furthermore, WT-metaD and US do not yield symmetric profiles, as one would expect for symmetric bilayers such as the one used in this study. For example, our WT-metaD calculations predict that the free energy is highest at $z \approx -0.7$ nm, and that the free energy difference to the upper membrane/water interface is 170 kJ/mol while that to the lower interface is only 50 kJ/mol. If we had calculated the free energy profile only for the $z \geq 0$ half-space and symmetrized it, we would have predicted a free energy maximum of 90 kJ/mol and would have lost the information about the asymmetry.  Such a profile would also have a sharp kink at $z=0$ nm, which would indicate a discontinuity in the mean force which is physically unlikely. This shows that mirroring the free energy profile at $z=0$ nm or symmetrizing it across the bilayer can be dangerous short-cuts that can yield misleading results. Visual inspection of the WT-metaD trajectory shows that phosphate groups from lipid headgroups interact strongly with the arginine and are pulled along as the solute is driven across the bilayer by the metadynamics algorithm. This displacement of lipid molecules due to strong interactions is likely the origin for the observed asymmetric free energy profile. It has been previously observed by Neale and co-workers who studied translocation of n-Propylguanidium across a DOPC membrane. During translocation of the solute, the membrane forms a depression that facilitates the continued interaction of the bilayer surface with the solute \citep{Neale2011}. 

US results show similar characteristics to results obtained with WT-metaD. The maximum of the barrier does not occur at $z=0$ nm and the free energy differences between the maximum and the upper and lower interfaces are different. The barrier height, however, is smaller than that predicted by WT-metaD. It is important to note that even sophisticated error analysis techniques such as bootstrapping cannot accurately quantify the discrepancy between the computed and the actual free energy profile. The obtained uncertainties are on the order of 15kJ/mol for our US simulations, which severely underestimates the actual error of the calculation. This is not surprising, as such error estimates can only use information from the sampled trajectories, and by nature have no knowledge of hitherto unsampled configurations.

Visual inspection of the simulation centered at $z=0$ nm shows that throughout the simulation arginine is in contact with lipid headgroups from the upper leaflet only. This persistent association with molecules from only one leaflet is an artifact of the initial configuration caused by insufficient equilibration. We constructed the initial configuration by pulling arginine from bulk water into the membrane using WT-metaD. Our observation suggests that even after 4 $\mu$s (100ns/window) of US simulation, the system still remembers its initial configuration and is therefore not equilibrated. This result is also consistent with previous studies of translocation of arginine side chain analogs across DOPC bilayers where sampling error persisted for 125 ns~\cite{Neale2011}.     

The UE simulation method provides additional information, not present in WT-metaD or US calculations, that aid in the detection of such convergence failures. In principle each trajectory should diffuse through the complete space of replicas due to the ongoing exchange between neighboring replicas. Fig.~\ref{fig:UEarginine}(a) visualizes the trajectories of 40 replicas, spanning the bilayer from $-2.0$ nm$<$ z $\leq$ 2.0 nm. Because exchange occurs frequently between most replicas, it is difficult to discern individual trajectories, and we highlight three specific ones to illustrate the motion of trajectories through replica space. 

There is, however, a well-defined and persistent gap in the exchange pattern, which is visible as a white line in Fig.~\ref{fig:UEarginine}(a). This gap indicates that the two neighboring replicas do not exchange configurations over long periods of simulation time. The location of this gap is shifting slowly on the timescale of the simulation. Its effect on the exchange process can be seen by the highlighted trajectory that starts in the replica centered at $z = 2$ nm: it diffuses freely though replica space, but when it hits the gap at approximately 260 ns of simulation time, it cannot cross to the other side of the gap; instead it is reflected back towards replicas centered at positive values of $z$. This gap separates replica space into two regions that do not exchange configurations with each other.

Visual inspection reveals that this gap separates configurations based on the association of the arginine solute with lipid headgroups from the two different leaflets. In replicas above the gap, the solute is in close contact with lipid headgroups from the upper leaflet (panels (c) and (d) of Fig.~\ref{fig:UEarginine}), whereas in trajectories below the gap it is in contact with either both or just the bottom leaflet (Fig.~\ref{fig:UEarginine}(e)). At approximately 280 ns the gap moves across the $z=0$ replica. After this jump, arginine is in close contact with lipid headgroups from both leaflets (Fig.~\ref{fig:UEarginine}(f)). Only after a long simulation time does the $z=0$ replica lose the memory that its initial configuration was chosen in a way that the solute had a higher affinity towards the upper leaflet.

Given the asymmetric profiles generated using WT-metaD and US, we hypothesized that the configuration shown in Fig.~\ref{fig:UEarginine}(f) is more representative for the equilibrium ensemble of the $z=0$ replica. We therefore performed a new UE calculation using this configuration as the starting point for replicas inside the bilayer. Fig.~\ref{fig:UEarginine}(b) shows the exchange patterns of the replicas for this second UE calculation. It no longer exhibits an apparent gap such as the one visible in Fig.~\ref{fig:UEarginine}(a), and replicas can explore the entire replica space. This UE simulation yields the symmetric free energy profile shown in Fig.~\ref{fig:Farginine}, with a maximum of  50 kJ/mol at the center of the bilayer ($z=0$) as one would expect from a converged free energy calculation. In this case, convergence is also indicated by the very small magnitude of the estimated uncertainty. Such a symmetric profile was also observed by Neale and coworkers, who calculated the free energy profile for translocation of an arginine side chain analog across a 1-palmitoyl-2-oleoyl-sn-glycero-3-phosphocholine (POPC) bilayer using Virtual Replica Exchange~\cite{Neale2013}. They attributed the symmetry of the free energy to better sampling allowed by the ability of replicas to sort themselves along the order parameter. The free energy we calculate is lower than previously reported values of arginine translocation across DOPC bilayers~\cite{MacCallum2011, Dorairaj2007}. These studies considered the side chain analog of arginine, while we study the entire amino acid residue, which is the likely reason for this discrepancy. All reported free energy barriers for the translocation of a single arginine residue or its analogs are significantly higher than those found in experimental~\cite{Wimley1996, Hessa2005} and simulation~\citep{Gumbart2011} studies of arginine insertion as part of a trans-membrane peptide.

To further test whether the success of the second UE calculation and the failure to converge of the WT-metaD and US calculations are related to an erroneous bias of the solute towards either side of the membrane we calculate the committor distribution $P(p_\text{B})$ of the $z=0$ ensemble generated by these three methods. The committor $p_\text{B}$ of a configuration is the probability that in a trajectory initiated from this configuration with randomized velocities, the solute reaches the lower membrane/water interface before it reaches the upper one. By definition, configuration with $p_\text{B} = 0.5$ are transition states. Intuitively one might expect that many (if not all) equilibrium configurations in which the solute is at the center of the bilayer have a committor value of 0.5, i.e., the solute is equally likely to go to the upper or the lower interface. In this case a histogram of $p_\text{B}$ values over an ensemble of such configurations should be sharply peaked at $p_\text{B} = 0.5$.

\begin{figure}[bt]
\includegraphics[width=\columnwidth]{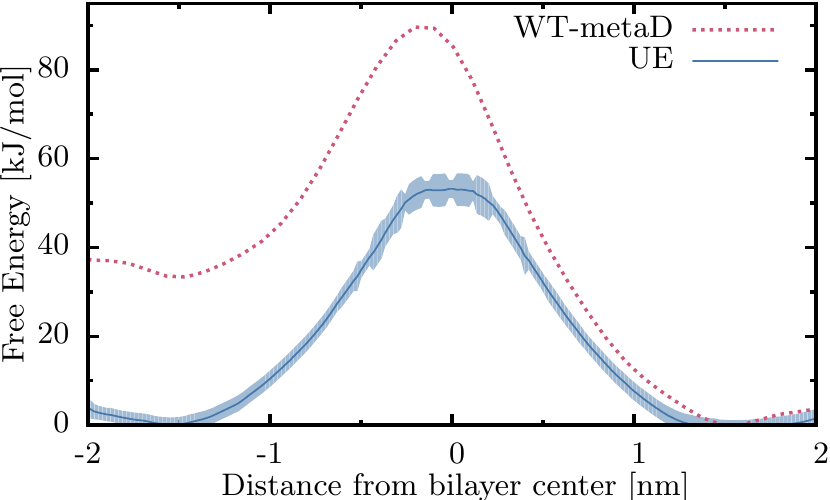}
\caption{\label{fig:Fsodium} Free Energy of sodium translocation as a function of distance from the bilayer center using WT-metaD and UE with shaded error bars for UE. The two lines differ: profiles generated using WT-metad is asymmetric and the maxima does not lie at $z=0$ nm. UE profile after extensive equilibration is symmetric with maximum at $z=0$ nm.}
\end{figure}

Details of the committor calculation are given in the Methods section, and results are shown in Figure~\ref{fig:committordistributions}. The committor histogram for the UE ensemble is peaked at 0.5 but the distributions for WT-metaD and US are not. The former indicates that the second UE calculation creates an unbiased ensemble of configurations that have no preference towards either leaflet. Configurations with $z=0$ as sampled by the WT-metaD and US methods, on the other hand, have a significant bias towards one leaflet over the other.  For the WT-metaD ensemble, 3 out of the 6 initial configurations we chose occurred while the arginine was crossing the bilayer from top to bottom, and all these configurations returned to the upper membrane/water interface in all four trajectories. In the remaining 3 initial configurations the solute crossed the membrane from bottom to top, all of those returned to the bottom leaflet when four trajectories were initiated from them. Together, this gives rise to the double-peaked histogram shown in Fig.~\ref{fig:committordistributions}. It shows that in a configuration in which the solute is at the bilayer center encountered along a WT-metaD trajectory, the solute has a strong tendency to return to the side from which it just came, an effect similar to hysteresis. All configurations sampled by the US algorithm, on the other hand, exhibited a strong bias towards the upper membrane/water interface, resulting in a histogram with a single peak at $p_\text{B} = 0$. The likely reason for this is that the initial condition for the $z=0$ window in the US calculation was taken from the WT-metaD trajectory where the arginine had been pulled into the membrane from above. The large peak at $p_\text{B} = 0$ shows that the system remembers this initial configuration even after 100 ns of simulation time.

\begin{figure*} [tb]
\includegraphics[width=\textwidth]{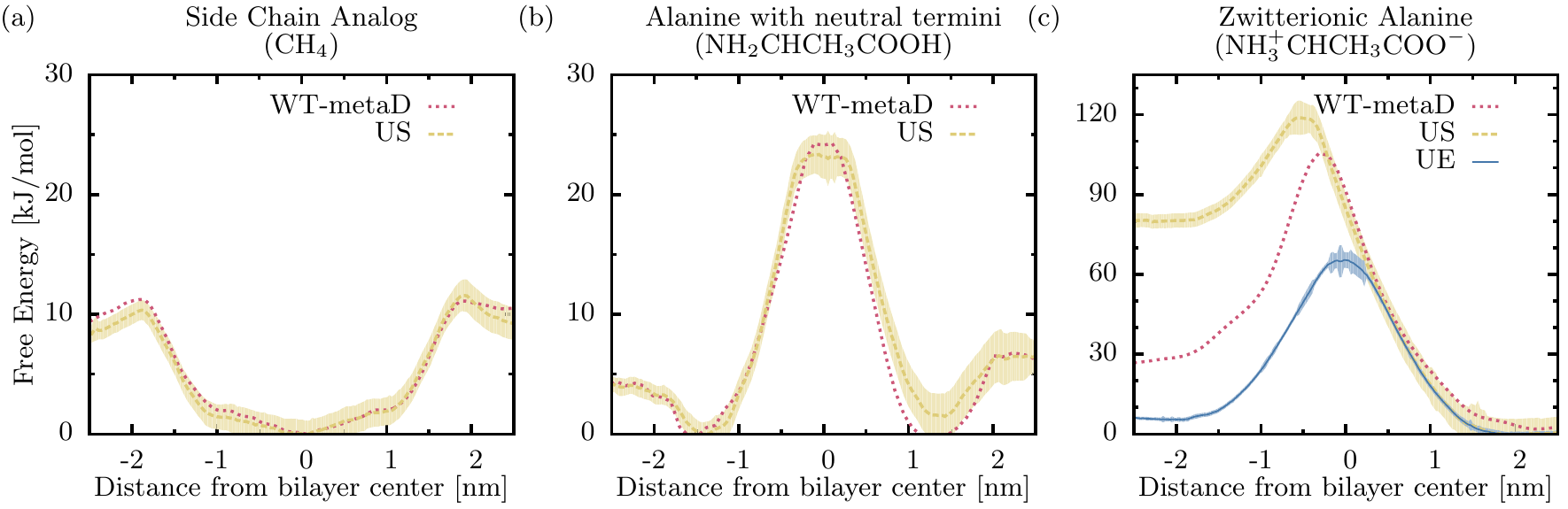}
\caption{\label{fig:Falanine} Free Energy of three forms of alanine translocation as a function of distance from the bilayer center with shaded error bars for US and UE. WT-metaD and US profiles are comparable for side chain analog of alanine (a) and alanine with neutral termini (b). For zwitterionic alanine (c), the three lines differ: profile generated using WT-metad and US are asymmetric and the maxima does not lie at $z=0$ nm. UE profile after extensive equilibration is symmetric with maximum at $z=0$ nm.}
\end{figure*}

Irrespective of the method used, capturing the transition state and generating the free energy profile for arginine translocation across a DOPC bilayer is challenging and requires significant simulation time. The UE method has the advantage that it provides a useful diagnostic tool to identify potential convergence issues: one can study the pattern of replica exchanges to check for unexpected behavior of the simulation, such as the gap shown in Fig.~\ref{fig:UEarginine}(a). This is an important advantage of this method over the others, even though it is not directly related to its sampling efficiency.

\subsection{Sodium Ion}

To test whether strong electrostatic interactions between zwitterionic lipid head groups and a charged solute are the reason for long relaxation timescales we performed WT-metaD and UE simulations of a single sodium ion translocating across the membrane. The resulting free energy profiles, shown in Figure~\ref{fig:Fsodium}, resemble those of the arginine calculations: that  obtained from WT-metaD is not symmetric across the bilayer, and has a maximum that is slightly offset from the bilayer center. Inspection of our WT-metaD simulation trajectory reveals that lipid molecules whose headgroups are in close contact with the ion are pulled along as the latter is driven across the membrane. US can yield similar, asymmetric free energy profiles for sodium translocation, as reported in the literature ~\citep{Wilson1996}. 

The replica exchange pattern of the UE simulation shows a persistent gap that slowly moves through replica space (Fig.~S1(a)), similar to that seen in the arginine calculations. This gap separates replicas based on which phospholipid headgroups are in close contact with the sodium ion. For example, in the initial condition of the $z=-0.2$ nm replica the sodium ion was surrounded by headgroups from upper leaflet lipids. After approximately 35 ns the gap crosses this replica, which from that point onward contains conformations in which headgroups from both leaflets are in contact with the ion. As before, starting a new UE calculation from those configurations generates an exchange pattern without a persistent gap (Fig.~S1(b)), and yields a symmetric free energy profile (Fig.~\ref{fig:Fsodium}). We find that the change in free energy of moving the ion from the membrane/water interface to the center of the bilayer is approximately 50 kJ/mol, which is surprisingly close to the value we obtained for arginine. A similar observation was made by Vorobyov and coworkers who found that translocation free energy barriers are similar for an arginine side chain analog and simple ions in 1,2-dipalmitoyl-sn-glycero-3-phosphocholine (DPPC) bilayers~\cite{Vorobyov2014}.

\subsection{Alanine}

Both arginine and sodium are cationic. To test whether the presence of a net charge on the solute is a prerequisite for the long equilibration timescales that we observe, we study the translocation free energy of three different forms of alanine. The first is its side chain analog, which is constructed by replacing the peptide backbone including the $\alpha$-carbon by a single hydrogen atom; this form is identical to a methane molecule. The second form of alanine is built using charge-neutral peptide termini NH$_2$ and COOH. The third form has the charged termini NH$_3^+$ and COO$^-$. All three forms have no net charge, but differ in the dipole moments of 0, 4.02 D, and 14.4 D, respectively, as calculated by the g{\_}dipoles tool of Gromacs. 

The side chain analog of alanine has neither a net charge nor a dipole moment, and therefore weak electrostatic interactions with phospholipid headgroups. As such one might expect that it does not stay in close contact with them as the solute is driven through the membrane, and that all methods should give comparable results even without extensive equilibration. Fig.~\ref{fig:Falanine}(a) shows that this is indeed the case: both WT-MetaD and US yield approximately symmetric free energy profiles that are comparable to each other within error bars. One key difference, which is unrelated to the used simulation method, is that the side chain analog is hydrophobic, and the free energy therefore has a minimum at the center of the bilayer.

For the second form of alanine with neutral termini, its dipole moment is sufficiently large to prefer the bulk water phase or the membrane/water interface region over the membrane interior, but not large enough to induce strong interactions with lipid headgroups. Therefore all methods yield converged results, as shown in Fig.~\ref{fig:Falanine}(b) for WT-MetaD and US.

Sampling the free energy profile of zwitterionic alanine, on the other hand, reveals similar behavior previously observed for the charged arginine and sodium solutes. Strong interactions between the solute and lipid headgroups cause the latter to be dragged along as the alanine is driven towards the interior of the bilayer by the WT-MetaD algorithm. These interactions do not relax on the simulation timescale, which leads to the asymmetric shapes of non-converged free energy profiles for WT-MetaD and US (Fig.~\ref{fig:Falanine}(c)). As was the case in the arginine calculation, the estimated error of the US profile is far too small, which is most likely due to insufficient sampling throughout the simulations. The existence of the electrostatic bottleneck can be detected by studying the exchange pattern of UE calculations, which exhibit a persistent gap indicating non-overlapping neighboring replicas~(Fig. S2(a)). Visual inspection of the trajectories reveals that replicas inside the bilayer initially contain configurations in which the solute is in close contact with lipid headgroups from only one leaflet, but transitions to configurations in which both leaflets' headgroups  are in contact with the solute once the gap has moved passed the replica. Using the latter conformations as the initial condition for a new UE calculation yields exchange patterns without persistent gaps (Fig. S2(b)) and a nearly symmetric, well-converged free energy profile (Fig.~\ref{fig:Falanine}(c)). We calculate a barrier height of 60 kJ/mol at the center of the bilayer.

These results show that long relaxation timescales and the resulting sampling problems are not unique to the translocation of charged solutes. Permeants with a sufficiently large dipole moment can exhibit the same behavior, and the analysis of UE exchange patterns can aid in the detection and resolution of such problems.

\subsection{Water}

\begin{figure} [bt]
\includegraphics[width=\columnwidth]{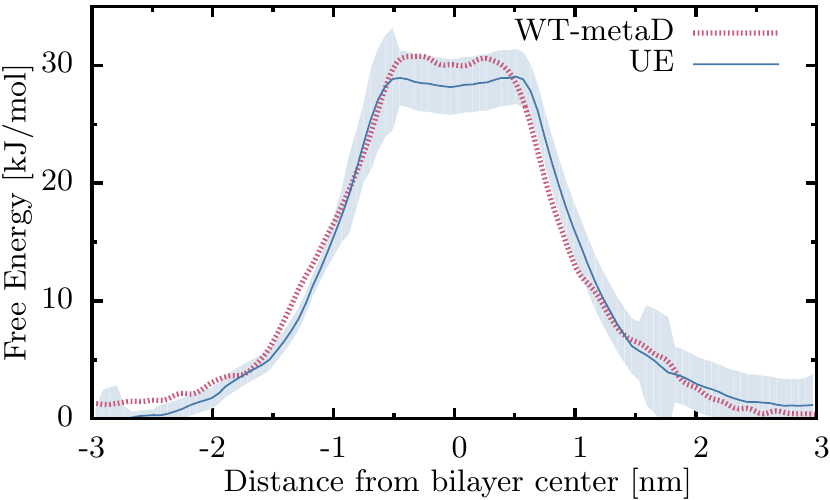}
\caption{\label{fig:Fwater} Free Energy of a single water molecule as a function of distance from the bilayer center using WT-metaD and UE. Uncertainties were computed using bootstrapping for UE results and are indicated by the shaded area. The two results are comparable, in particular both free energy profiles are symmetric with respect to the bilayer center.}
\end{figure}

Finally we calculate the free energy profile of translocation of a single water molecule across the bilayer. Within the SPC model used in our simulations a water molecule has a dipole moment of 2.27 D, which based on our previous analysis of alanine we expect is sufficiently small to avoid sampling difficulties due to long relaxation timescales. This is verified by the lack of a persistent gap in the UE exchange pattern (Fig.~S3) and the convergence of both WT-metaD and UE calculations to symmetric free energy profiles within error bars(Fig.~\ref{fig:Fwater}). We find a barrier height of 27 kJ/mol at the bilayer center, which is in good agreement with previous studies of water translocation through PC bilayers using the US method~\cite{Sapay2009, Orsi}. Visual inspection of the WT-metaD trajectory confirms that no lipid headgroups are dragged along the water molecule as it is driven through the membrane.

\section{\label{sec:conslusions}Discussion}

Our results demonstrate that the calculation of accurate translocation free energy profiles of solute motion across lipid bilayers can require long simulation times and care in the choice of initial conditions for all tested simulation methods. This is the case for the positively charged solutes arginine and sodium as well as the strongly dipolar zwitterionic alanine. In these cases strong electrostatic interactions between the solute and phospholipid headgroups create a long-lived affinity of the solute to one bilayer leaflet even when the solute is in the center of the bilayer that takes hundreds of nanoseconds to relax. Similar timescales related to membrane reorganization have been observed in simulations of peptides interacting with membranes~\cite{Yesylevskyy2009,Neale2014, Bennett2016}. Small, uncharged solutes with no (side chain analog of alanine) or small (neutral alanine, water) dipole moments do not exhibit this dynamical bottleneck.

Neither of the three tested simulation methods can intrinsically accelerate this relaxation process. However, replica exchange umbrella sampling (UE) has a desirable feature lacking both in umbrella sampling (US) and well-tempered metadynamics (WT-metaD): by analyzing the exchange pattern between replicas one can easily detect sampling deficiencies caused by mismatched configurations in neighboring replicas. These manifest themselves as persistent gaps in the exchange pattern that hinders efficient mixing in replica space. Movement or disappearance of this gap can aid in choosing initial conditions for subsequent simulations that do not suffer from this bottleneck. The importance of even exchanges between all replicas has been previously emphasized by Neale and coworkers, who developed a novel metric, the transmission factor, to quantify the diffusivity of replicas~\cite{Neale2013}.

In all cases it is advisable to compute the free energy profile for translocation across the entire lipid bilayer. Calculating the profile only for one half-space and then mirroring it to obtain the other is sufficient in principle and minimizes computing cost, but in practice leads to the loss of valuable information: for symmetric membranes, an asymmetric free energy profile is an indicator that the calculations are not yet converged. Similarly, symmetrizing a free energy profile obtained for the entire membrane should not be done blindly, as it might lead to an unwarranted confidence in the result.

In this work he have focused only on a single collective variable:  the normal distance between the bilayer center and the solute. This is by far the most widely used order parameter in free energy calculations of solute-bilayer interactions. Choosing different (or additional) collective variables for the biased sampling schemes might alleviate the convergence issues we encountered. However, the selection of the optimal order parameter in itself is a challenging problem and beyond the scope of this paper~\citep{Lin2015, Hinner2009, Wang2014}. 

Another field of ongoing research is the further refinement of molecular mechanics force fields to better describe the system at hand. It has been shown that both structural details of simulated bilayers and thermodynamic properties of membrane translocation vary among commonly used force fields~\citep{Piggot2012, Sun2015}. However, details of the force field selection should not alter our primary conclusion because the described sampling and convergence issues are rooted in basic physical properties, specifically the electrostatic interaction between the solute and the phospholipid headgroups.   

Judging the convergence of free energy calculation is not an easy task. Because common error analysis methods may not always reveal characteristics of a non-converged free energy profile, these calculations need additional verification. As useful guidance we propose the following diagnostic checks:
\begin{itemize}[noitemsep]
	\item[i.] Is the free energy profile symmetric with respect to the bilayer center? This is a strict requirement for symmetric membranes.
	\item[ii.] Does the free energy profile exhibit a plateau region in the center of the bilayer? Mirrored or symmetrized free energy profiles often exhibit a kink at $z=0$. This could be a warning sign for insufficient convergence, as such a kink would imply a discontinuity in the mean force, which is unlikely to be physically meaningful.
	\item[iii.] When performing an UE calculation with initial conditions taken from simulation trajectories at varying $z$-positions of the solute, is there an apparent gap in the resulting exchange pattern? The presence of such a gap indicates the existence of barriers in replica space that will impede proper sampling.
	\item[iv.] For solutes that are most likely found in bulk water or at the membrane-water interface, do the sampled configurations at the center of the bilayer exhibit a bias towards one side of the membrane over the other? For symmetric membranes it is likely that many configurations of the $z=0$ ensemble are transition states, i.e., they have equal probability of evolving towards states in which the solute is on either side of the membrane. Calculating the committor distribution function is an effective if time-consuming way to test this property: if it is peaked at $p_\text{B} = 0.5$ then one indeed samples the transition state ensemble, which is not the case if it is peaked at $p_\text{B} = 0$ and/or $p_\text{B} = 1$.
\end{itemize}

None of the three tested methods has an intrinsic advantage when comparing the amount of sampling that can be obtained for a fixed amount of computing time. They differ, however, in how computational resources are allocated. WT-MetaD calculations are typically performed as a single, long trajectory. This trajectory can be run on multiple computing nodes in parallel, but the scaling efficiency is limited by the size of the system. Other extensions of the metadynamics method, such as Parallel Tempering Metadynamics, allow the parallelization of multiple metadynamics trajectories in a single calculation~\cite{Bussi2006}. US, as frequently used and described here, involves performing multiple independent simulations, one for each biasing window. This can be done either in series or in parallel depending on the resources available, which provides the most flexibility. In contrast, general UE requires that simulations of all replicas are running in parallel, which necessitates the availability of many computing cores at the same time. However, there are variants of the Replica Exchange algorithm, such as Virtual Replica Exchange (VREX)~\citep{Rauscher2009} and Serial Replica Exchange~\citep{Hagen2007} that circumvent the need for synchronous simulations.  

In summary, we believe that the ability to judge convergence by examining the exchange patterns makes UE an excellent choice if the computing requirements can be satisfied. It yields robust free energy profiles, which are essential for gaining a deeper understanding of membrane translocation processes and for predicting permeation rates.
 
\section*{Acknowledgments}

We thank Chris Neale for helpful discussions. This work was facilitated though the use of advanced computational, storage, and networking infrastructure provided by the Hyak supercomputer system at the University of Washington. We thank the Royalty Research Fund for partial financial support for this work through Grant 65-4785. 

\bibliography{paper}

\clearpage
\onecolumngrid
\setcounter{page}{1}
\begin{center}
Supplemental Information 

\Large \bfseries Free Energy Calculations of Membrane Permeation: \\ Challenges due to Strong Headgroup--Solute Interactions
\end{center}

\setcounter{figure}{0}
\setcounter{table}{0}

\makeatletter 
\renewcommand{\thetable}{S\@arabic\c@table}
\makeatother

\begin{table}[h]
\caption{\textbf{Metadynamics Parameters}}

\begin{tabular} {| l | l | l | l | l |} 
\hline
\textbf{Solute}             		& Height(kJ/mol) & Width(nm) & Time of gaussian addition(ns) & Biasfactor \\\hline
Zwitterionic arginine 				& 15 & 0.3 & 3  & 140\\ \hline
Sodium 					& 5.0 & 0.3 & 3 & 75.0   \\ \hline 
Side chain analog of alanine         & 0.1 & 0.1 & 5 & 2.00  \\ \hline
Alanine with neutral termini    	& 10 & 0.3 & 3 & 35.0   \\ \hline
Zwitterionic alanine        & 15 & 0.3 & 3 & 140   \\ \hline
Water 					& 2.2 & 0.1 & 3 & 10.0    \\ \hline

\end{tabular}
\end{table}

\begin{threeparttable}[h]
\caption{\textbf{Lengths of Simulation Trajectories}}

  \begin{tabular}{|l|l|l|l|l|l|}
    \hline
    \multirow{2}{*}{\textbf{Solute}} &
      \multicolumn{2}{c}{\textbf{US}} &
      \multicolumn{2}{c}{\textbf{UE}} &
      \multirow{2}{*}{\textbf{WT-metad} ($\mu$s)}\\
& length/window (ns) & total ($\mu$s) & length/replica (ns) & total $^*$ ($\mu$s)  &  \\
  
    \hline
Zwitterionic arginine 	& 100 & 4.00 & 166  & 6.64 & 1.00\\ \hline
Sodium 					& - & - & 27 & 1.08 & 1.08   \\ \hline 
Side chain analog of alanine         & 20 & 1.00 & - & - & 1.00  \\ \hline
Alanine with neutral termini     	& 20 & 1.00 & - & - & 1.00   \\ \hline
Zwitterionic alanine        & 55 & 2.75 & 55 & 2.20 & 1.40   \\ \hline
Water 					& - & - & 8 & 0.32 & 0.42    \\ \hline

\end{tabular}
\begin{tablenotes}
       \item $^*$ {\small UE total simulation time only represent simualtions that span the membrane between $-2.0$nm$<z\leq 2.$nm. We ran additional simulation around $z-\leq2.0$ nm and $z>2.0$ nm to investigate if solute configuration changes in bulk water phase.}
    \end{tablenotes}
\end{threeparttable}

\makeatletter 
\renewcommand{\thefigure}{S\@arabic\c@figure}
\makeatother

\begin{figure} [h]

\includegraphics[width=\columnwidth]{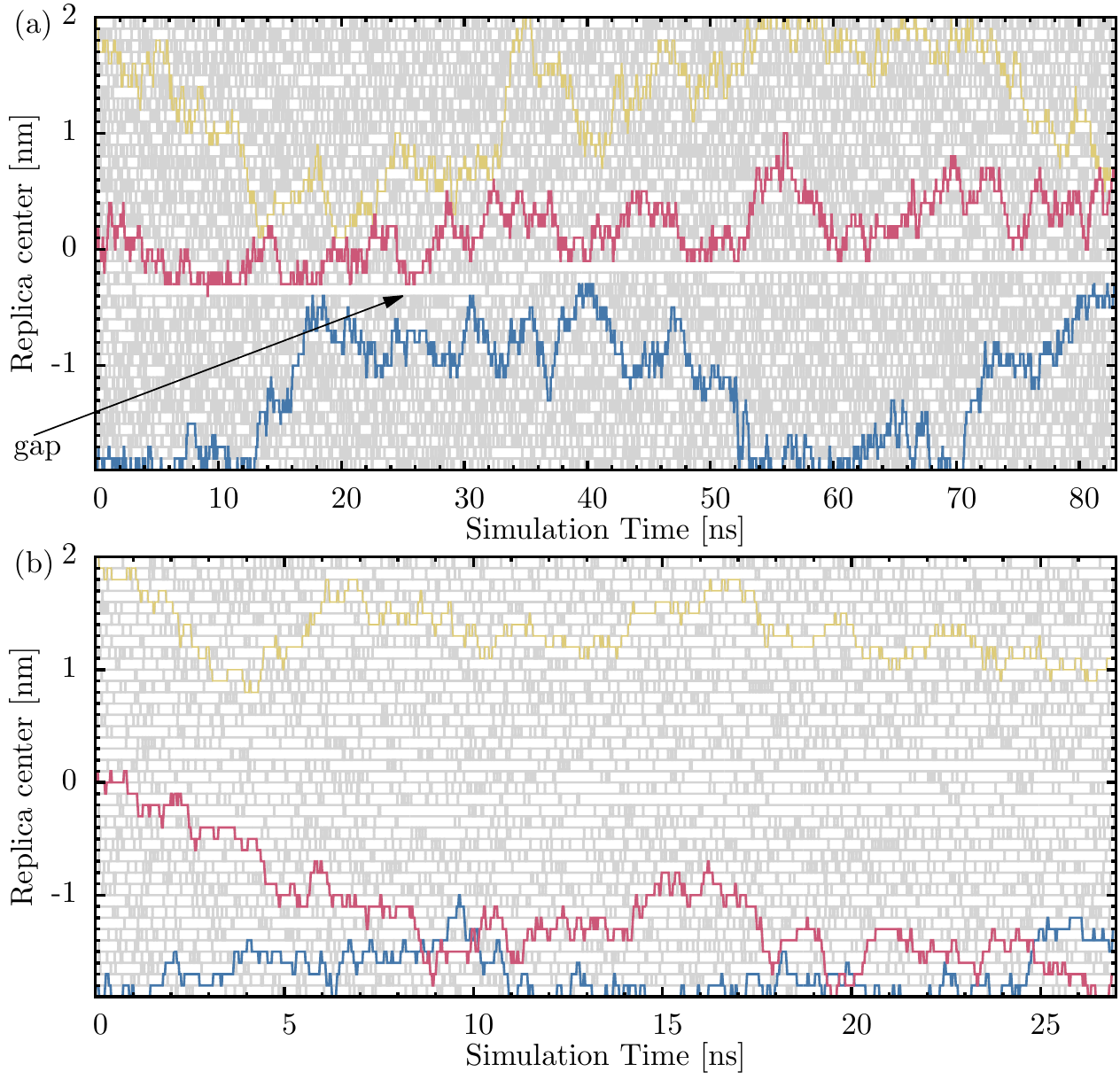}
\caption{\label{Supp Fig. 1} \footnotesize UE results for sodium. (a) Exchange pattern lines for 40 replicas. Replicas starting at $z=2.0$ nm, $0$ nm and $-1.9$ nm are highlighted. A gap in the exchange pattern (indicated by the arrow) is clearly visible. Using the final configuration of $z=0$ replica as the initial condition for the center replicas in a new UE simulation (b), this gap is no longer present, indicating even exchanges between all 40 replicas.}
\end{figure}

\begin{figure} [h]
\includegraphics[width=\columnwidth]{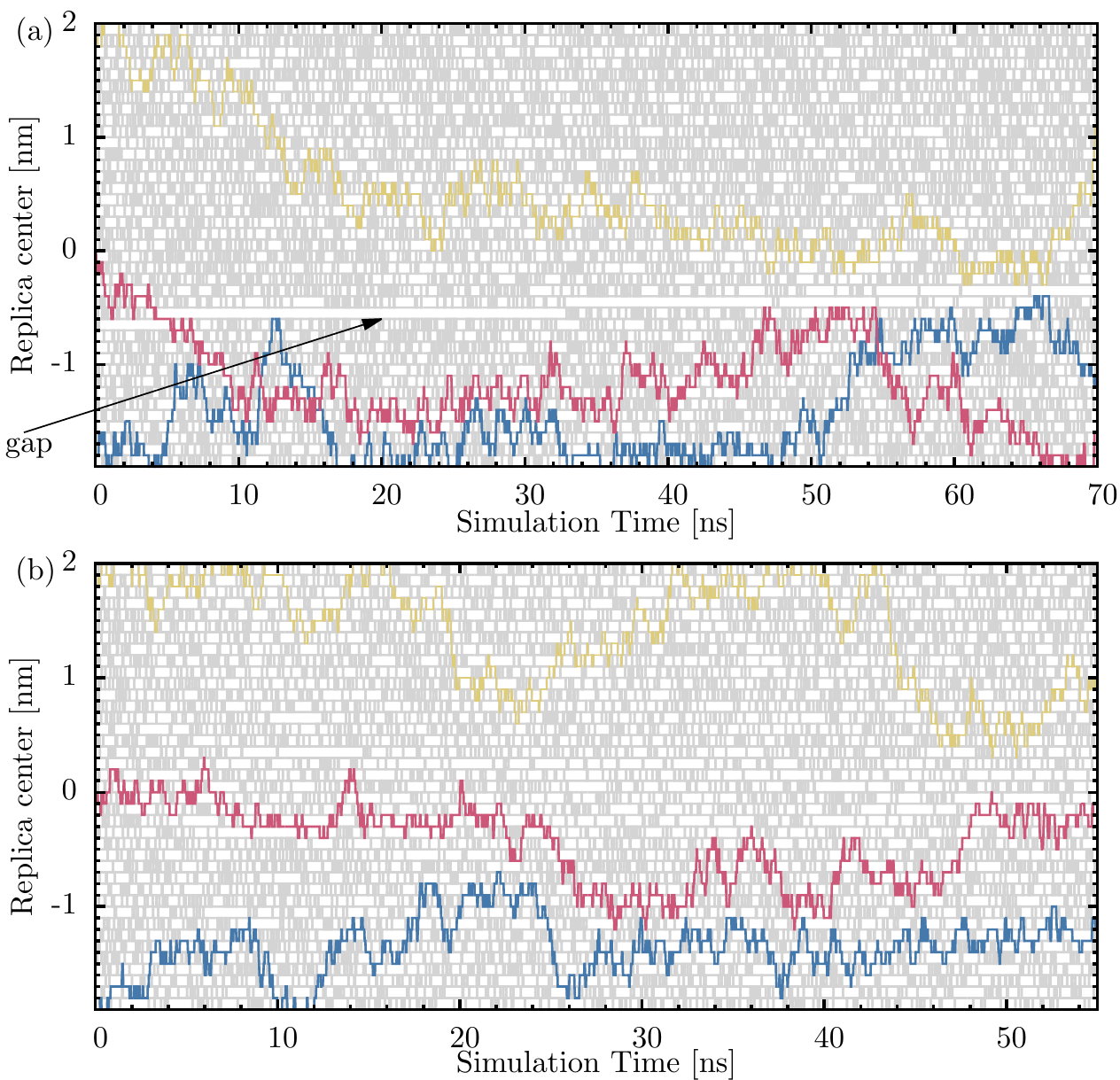}
\caption{\label{Supp Fig. 2} \footnotesize UE results for zwitterionic form of alanine. (a) Exchange pattern lines for 40 replicas. Replicas starting at $z=2.0$ nm, $0$ nm and $-1.9$ nm highlighted. A gap in the exchange pattern (indicated by the arrow) is clearly visible. Using the final configuration of $z=0$ replica as the initial condition for the center replicas in a new UE simulation (b), this gap is no longer present, indicating even exchanges between all 40 replicas.}
\end{figure}

\begin{figure} [h]
\includegraphics[width=\columnwidth]{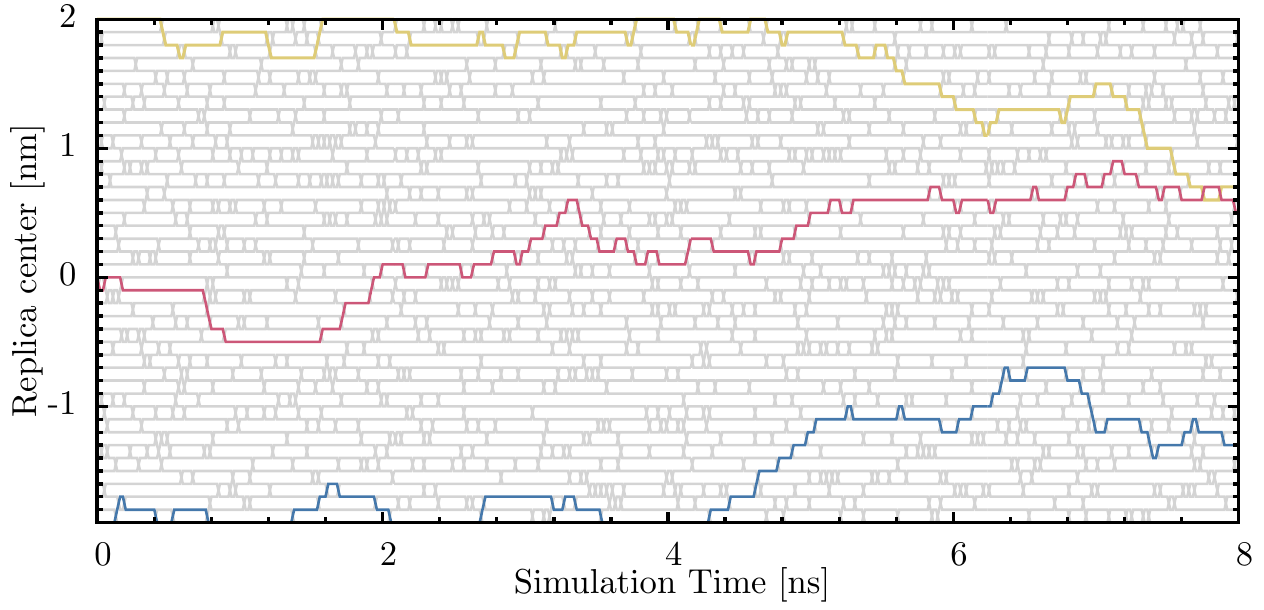}
\caption{\label{Supp Fig. 3} \footnotesize UE results for water. Exchange pattern lines for 40 replicas. Replicas starting at $z=2.0$ nm, $0$ nm and $-1.9$ nm highlighted. No gap in the exchange pattern exists indicating an even exchanges between all 40 replicas.} 
\end{figure}

\end{document}